\documentclass[%
 reprint,
 amsmath,amssymb,
 aps,
 prl,
]{revtex4-1}

\usepackage{graphicx}% Include figure files
\usepackage{dcolumn}% Align table columns on decimal point
\usepackage{bm}% bold math
\begin{document}

\title{Alchemical perturbation density functional theory (APDFT)}
%for rapid yet accurate quantum property estimates}% Force line breaks with \\

\author{Guido Falk von Rudorff}
\affiliation{Institute of Physical Chemistry and National Center for Computational Design and Discovery of Novel Materials (MARVEL), Department of Chemistry, University of Basel, Klingelbergstrasse 80, CH-4056 Basel, Switzerland}

\author{O. Anatole von Lilienfeld}
\email{anatole.vonlilienfeld@unibas.ch}
\affiliation{Institute of Physical Chemistry and National Center for Computational Design and Discovery of Novel Materials (MARVEL), Department of Chemistry, University of Basel, Klingelbergstrasse 80, CH-4056 Basel, Switzerland}

\date{\today}

\begin{abstract}
%Perhaps the abstract could be made more general and clear though.
%I did not catch the major advance upon reading the abstract.
%our numerical results suggest that 
%for perturbed reference electron densities evaluated at several levels of theory (LDA, GGA, hybrid, and CCSD) 
%The electron density of {\em any} given iso-electronic target system is estimated using %approximated within a Taylor 
%The approach works best when reference and target share the same geometry. 
%The approach is exemplified for the toy system He and H$_2$, as well as for diatomics N$_2$, CO, and BF, 
%and our calculations indicate rapid convergence with perturbation order.
We introduce an orbital free electron density functional approximation based on alchemical perturbation theory. 
Given convergent perturbations of a suitable reference system, 
the accuracy of popular self-consistent Kohn-Sham density functional estimates of properties
of new molecules can be systematically surpassed---at negligible cost. 
The associated energy functional is an approximation to the integrated energy derivative, 
requiring only perturbed reference electron densities: No self-consistent field equations are necessary to estimate
energies and electron densities. 
Electronic ground state properties considered include covalent bonding potentials, atomic forces, as well as dipole and quadropole moments. 
\end{abstract}

%\pacs{Valid PACS appear here}
\maketitle

%\tableofcontents
%\section{Introduction}
With the success of electronic structure methods in the materials, chemical, and biological sciences, 
the need for ever more accurate yet ever computationally more affordable methods grew. Approaches like the Harris functional\cite{Harris1985,Averill1990} tried to employ an approximate density rather than a fully self-consistent one by following the Kohn-Sham scheme\cite{Kohn1965} for one step only. While the resulting energies have been shown to be of acceptable accuracy for bulk crystals\cite{Polatoglou1988,Read1989}, the difference in density however is quite significant\cite{Finnis1990} and the energies of the Harris functional are neither upper nor lower bounds to the self-consistent energy\cite{Robertson1991,Zaremba1990}. This has been attributed to the non-variational approach and subsequently addressed by treating the approximate wavefunction as perturbation to the true wavefunction\cite{Benoit2001,Zhu2004}. In line of applications however, this concept faced technical difficulties depending on the exchange-correlation functional employed\cite{Zhou2008} and was found to depend strongly\cite{Zhou2008} of the quality of the approximate density which often has been obtained by superimposing self-consistent fragment densities as suggested by Harris. Nevertheless, the Harris approach to employ (perturbative) approximate densities has been useful in improving convergence\cite{Zhou2008} or in deriving kinetic energy functionals in the context of orbital-free DFT\cite{Zhou2006}.
Other approaches were introduced by Foldy and Wilson, Reif, Frost, and Daza~\cite{Foldy1951,Wilson1962,Frost1962,Reif1984,Daza2005}. The transition functional method of Nagy\cite{Nagy1996} allows to calculate energy differences of two molecules if both their electron densities are known. Similar approaches have helped addressing the hard problem of a reliable kinetic energy expression in the context of orbital-free DFT\cite{Nagy2008,Nagy2011,Levaemaeki2015}.

If and only if densities change smoothly along iso-electronic integration path, the mean value theorem mandates that evaluating the integrand once for one (unknown) point on the integration path is sufficient to obtain an accurate energy\cite{Levy1978}. 
Based on scaling nuclear charges, a relation for the ground state energy as a function of the electrostatic potential at the nuclei was given\cite{Politzer1974}. 
It has also been suggested to expand the total energy in polynomials of the nuclear charges\cite{Linderberg1961}. This expansion converges quickly for small systems\cite{Stillinger1966} and can treat the nuclei-electrons and electron-electron interactions\cite{Politzer1978}. Despite the parametrization, the model was used to show conceptionally that the electron-electron interaction energy is limited in isoelectronic molecular series\cite{Castro1983} and to propose bounds on neutral atom energies\cite{Levy1980}.

More recently, alchemical perturbations in the spirit of Foldy and Wilson gained traction. 
In analogy to the well-established adiabatic coupling in the context of e.\,g. free energy calculations, the electronic Hamiltonians of two (isoelectronic) systems are adiabatically coupled via an arbitrary path described by a single mixing parameter, similar to the integration paths between molecules in the Wilson scheme\cite{Kryachko1984,Levy1978}. From the perspective of one of the molecular endpoints, the change in nuclear charges then can be considered to be a perturbation\cite{anatole-prl2005,anatole-jcp2009-2}. Although this perturbation is by no means small, the approach has been successful in screening of alkali halide crystals\cite{Solovyeva2016}, estimating the chemical potential of binary mixtures\cite{Alfe2000}, calculating bond potentials\cite{Samuel-CHIMIA2014,Samuel-JCP2016},  estimating energies, structures and volume in solid metals\cite{MoritzBaben-JCP2016}, band-structures in III-V semiconductors\cite{Samuel2018bandgaps}, predicting reaction barriers and molecular adsorption on metals\cite{Sheppard2010,saravanan2017alchemical}, predicting changes in adsorption energy of water on graphene due to BN doping\cite{Yasmine-JCP2017}, calculating
higher order energy derivatives\cite{LesiukHigherOrderAlchemy2012}, exploring chemical space\cite{CCSexploration_balawender2013}, predicting BN doped C$_{60}$\cite{balawender2018exploring}, 
or probing the non-local nature of the electron density~\cite{StijnPNAS2017}. 
By contrast, in this letter we describe the application of alchemical perturbation theory to the electron density, resulting in an orbital-free alchemical perturbation density functional theory (APDFT) formulation. 
Given a single reference electron density and energy, accurate electron densities and energies of iso-electronic query systems with identical nuclear positions are obtained at negligible computational cost. 
% discussion figure horder
\begin{figure}
\includegraphics[width=\columnwidth]{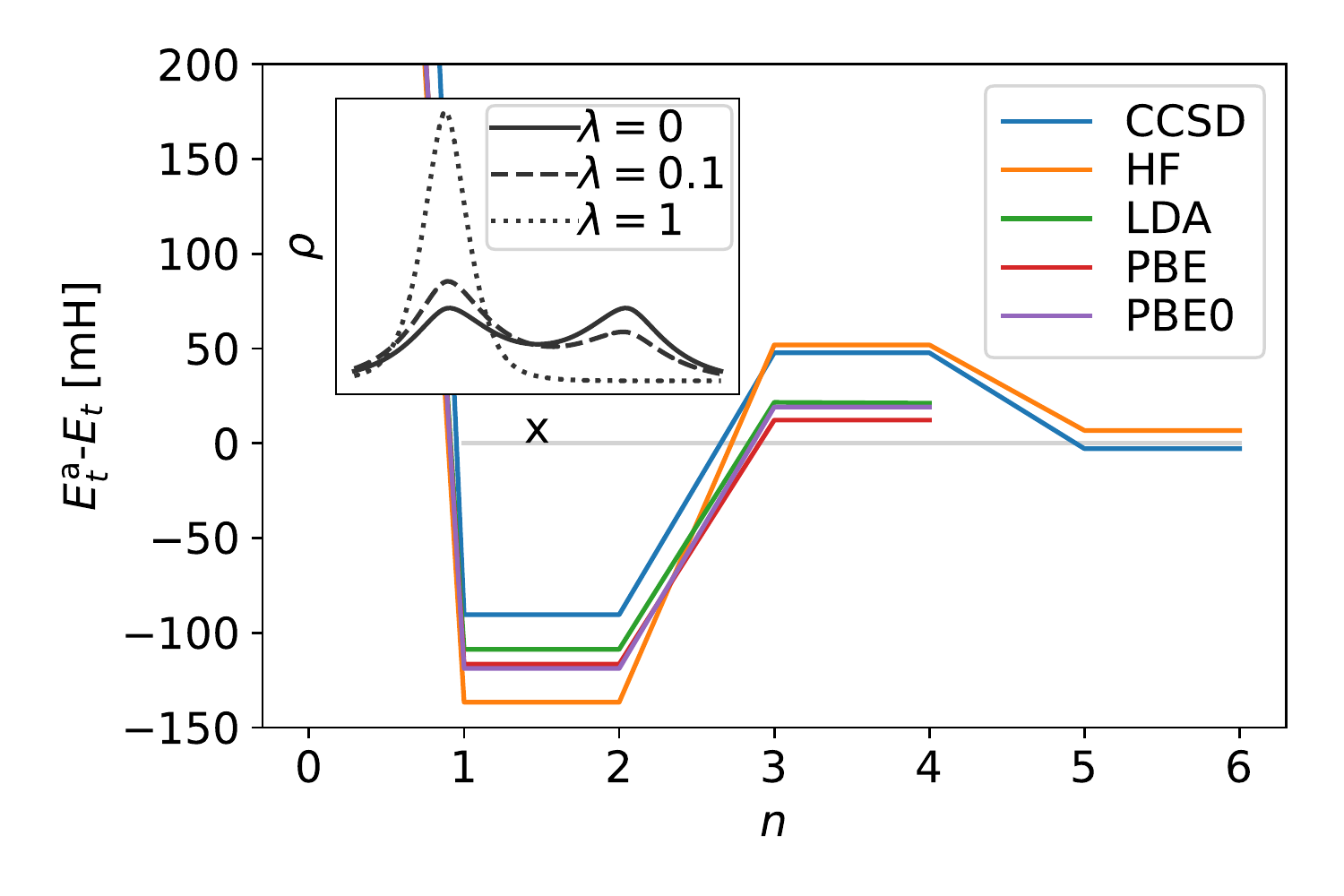}
\caption{Electronic energy error of He  as a function of expansion order $n$ (Eq.~\ref{eqn:totalenergy}) evaluated using HF and various DFA. 
CCSD values from (Eq.~\ref{eqn:hellmann}) applied to a polynomial fit of $E(\lambda)$.
The reference system is H$_\text{2}$ at 1\AA\; interatomic distance with def2-TZVP basis set.
Inset shows the HF/def2-TZVP electron density profile along the molecular axis for three $\lambda$ values.\label{fig:horder}}
\end{figure}

%\section{\label{sec:methods}Methods}
The overall goal is to calculate the electronic energy and the electron density of some target molecule if the total electronic energy and the electron density and the derivatives thereof are known for some reference molecule that is identical in geometry, but may differ in atomic composition. This is achieved via alchemical perturbation, i.e.~typically coupling the two involved electronic molecular Hamiltonians via a linear mixing parameter $\lambda$ as 
$
\hat H(\lambda)\equiv \lambda \hat H_\text{t} + (1-\lambda)\hat H_\text{r}$.
The resulting energy for a system can be expanded in a Taylor series around the reference molecule (i.e. $\lambda=0$) 
$
E_\text{t} \equiv E(\lambda=1) =\left.\sum_{n=0}^{\infty }\partial^n_\lambda \left \langle \psi_\lambda \left| { \hat H(0)} \right| \psi_\lambda \right\rangle/n!\right\rvert_{\lambda=0}
$
which can be expressed as 
\begin{align}
E_\text{t} &= \sum _{n=0}^{\infty }{\frac {1}{n!} \left.\frac{\partial^n E(0)}{\partial \lambda^n}\right\rvert_{\lambda=0}}
\; =\; E_\text{r} + \sum _{n=1}^{\infty }{\frac {1}{n!} \left.\frac{\partial^n E(0)}{\partial \lambda^n}\right\rvert_{\lambda=0}}\label{eqn:hellmann}
\end{align}
According to the Hellmann-Feynman theorem\cite{Feynman1939}, 
the first order partial derivative is the difference in external potential $v$ acting on any pair of iso-electronic molecular Hamiltonians~\cite{anatole-jcp2009-2},
\begin{align}
\frac{\partial E(\lambda)}{\partial \lambda} &= \langle \psi_\lambda | \hat H_\text{t} - \hat H_\text{r} | \psi_\lambda\rangle  \; = \;
%\underbrace{E^\text{nn}_\text{t} - E^\text{nn}_\text{r}}_{\equiv \Delta E^\text{nn}} + 
\int d\mathbf{r} \underbrace{(v_\text{t}(\mathbf{r}) - v_{\text{r}}(\mathbf{r}))}_{\equiv \Delta v}\rho_\lambda (\mathbf{r}),
\end{align}
%where $E^\text{nn}$ is the nuclear-nuclear energy contribution that can be different for the reference and target molecule due to their difference in atomic composition. 
and higher order partial derivatives correspondingly from further differentiation
$ %\begin{align}
{\partial^n_\lambda E(\lambda)} = \int d\mathbf{r} \Delta v \partial^{(n-1)}_\lambda \rho_\lambda$. %\label{eqn:deltav}
Insertion into Eq.~\ref{eqn:hellmann} gives for the change in energy,
\begin{align}
E_\text{t} - E_\text{r} &= \sum _{n=1}^{\infty }{\frac {1}{n!} \int d\mathbf{r} \Delta v\left.\frac{\partial^{n-1} \rho_\lambda}{\partial \lambda^{n-1}}\right\rvert_{\lambda=0}}
\end{align}
where $\partial^0_\lambda \rho = \rho$. 
This integral can be restricted to the finite volume $\Omega$, either because both $\rho_\text{t}(\mathbf{r})$ and $\rho_\text{r}(\mathbf{r})$ become zero far from the nuclei or because periodic boundary conditions require a finite unit cell. Further assuming uniform convergence of the sum allows to switch the sum and the proper integral:
\begin{align}
E_\text{t} - E_\text{r}  & = \int_\Omega d\mathbf{r} \Delta v\underbrace{\sum _{n=1}^{\infty }{\frac {1}{n!} \left.\frac{\partial^{n-1} \rho_\lambda}{\partial \lambda^{n-1}}\right\rvert_{\lambda=0}}}_{\equiv \tilde \rho} \; \nonumber \\
&= \; \int_\Omega d\mathbf{r} \Delta v(\mathbf{r}) \tilde \rho(\mathbf{r}) \label{eqn:totalenergy}
\end{align}
The sum builds a new shadow electron density, $\tilde\rho$ which we can understand using integration, 
\begin{align}
E_\text{t} - E_\text{r}  & = \int_0^1 d\lambda \frac{\partial E}{\partial \lambda } \;\; = \;\;
%\int d\lambda \int_\Omega d\mathbf{r} \Delta v({\bf r}) \rho_\lambda \nonumber\\
 \int_\Omega d\mathbf{r} \Delta v({\bf r}) \int_0^1 d\lambda \rho_\lambda({\bf r}).
 \label{eq:TI}
\end{align}
Expansion of $\rho_\lambda$ as a Taylor series in $\lambda$
\begin{align}
\rho_\text{t} \equiv \rho(\lambda=1) = \rho_\text{r} + \sum _{n=1}^{\infty }{\frac {1}{n!} \frac{\partial^n \rho(0)}{\partial \lambda^n}}\label{eqn:rhotarget}
\end{align}
recovers exactly the expression for $\tilde{\rho}$.
Thus, $\tilde \rho$ is neither the density of the reference nor the target. It rather corresponds to the lambda-averaged density. 
Already in 1978, Levy has shown the existence of such a density that allows calculation of all energy contributions via Eq.~\ref{eqn:totalenergy}\cite{Levy1978}. While Levy approximated $\tilde{\rho}$ by the average of $\rho_r$ and $\rho_t$, 
we rather focus on its approximation through the Taylor expansion in $\rho_r$ which is crucial for rendering the 
computational investment constant and independent of target system.

\begin{figure*}
\includegraphics[width=\textwidth,clip,trim=1cm 2cm 2.0cm 2.5cm]{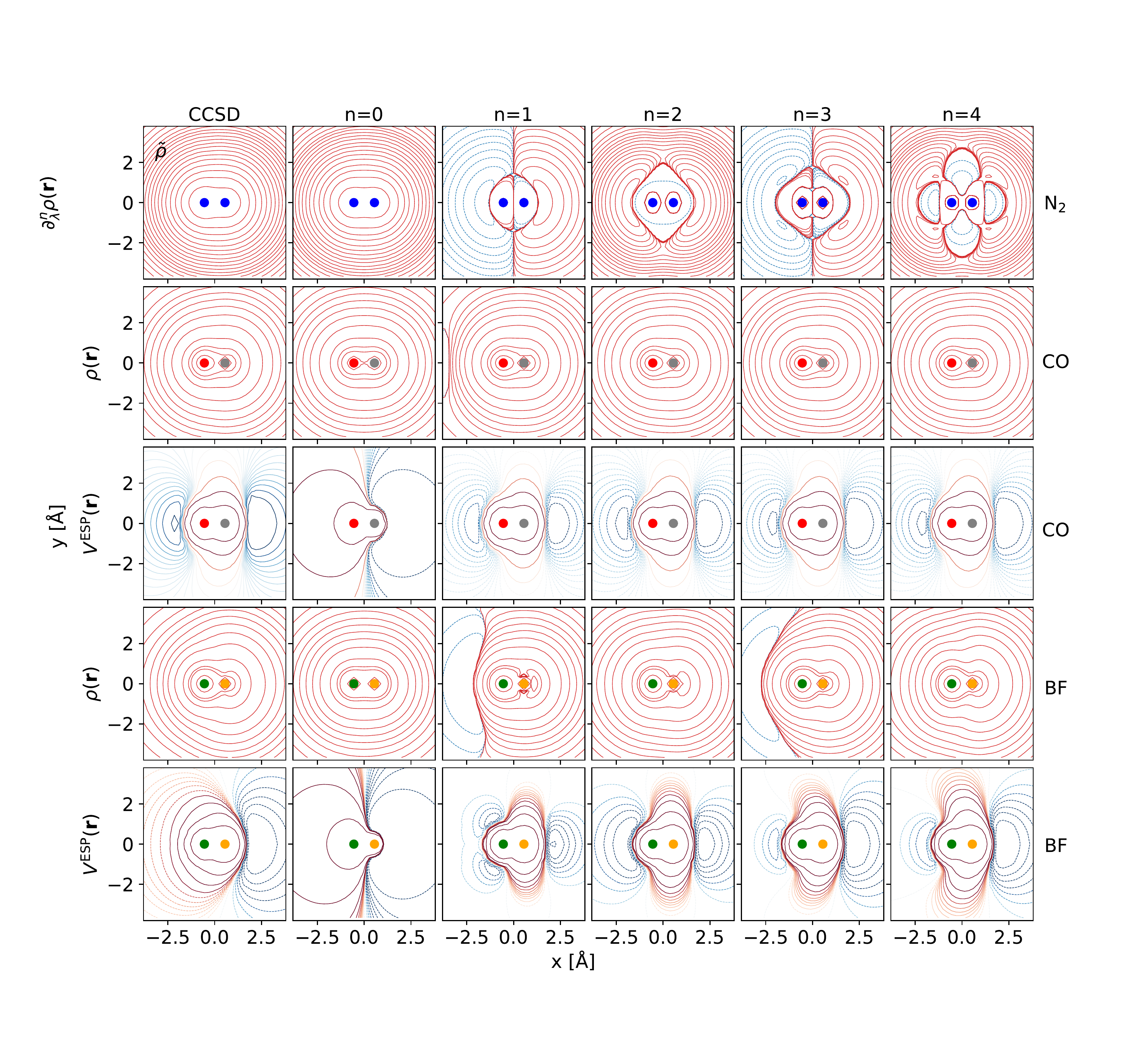}
\caption{
Top row: Left hand panel shows shadow-density $\tilde{\rho}$ of N$_2$ (from Eq.~\ref{eqn:totalenergy}), 
obtained from the individual (anti-)symmetrized electron density derivatives (other panels in row one) 
of the different orders of perturbations towards CO/BF. CO and BF total electron densities (second and fourth row) and the electrostatic potential (third and fifth row) in the bond plane 
are compared to CCSD/def2-TZVP (left hand column). 
Electron density and electrostatic potential converge faster for CO than for BF. 
Contour levels for the electron density and electrostatic potential are shared within the respective row and are derived from evenly spaced percentiles of the electron density. 
The density derivatives have independent contour levels. All red contour lines denote positive values, blue contour lines denote negative ones. Nuclei are colored according to their respective element. Multipole convergence shown in Table~\ref{tab:ccsd}.
\label{fig:ccsdesp}}
\end{figure*}

For N$_2$, we exemplify $\tilde \rho$ in Fig.~(\ref{fig:ccsdesp}).
Note also that Eq.~(\ref{eq:TI}) implies convergence in $\lambda$ as long as $\partial_\lambda E$ does not diverge.
Using Kato's cusp condition one can also demonstrate convergence for free atoms (see SI).
%Moreoever, the mean value theorem mandates that there is at least one density that gives the correct energy difference between reference and target molecule\cite{Levy1978} if densities vary smoothly in $\lambda$. 
While we show convergence for hydrogenic atoms and free atoms (see SI), we cannot offer a mathematically rigorous proof of convergence for all systems. It is likely, however, that the electron density can be described by an analytic function, i.e. a function with a converging Taylor series, since it is common in quantum chemistry calculations to approximate the electron density with Gaussian functions, both in the context of Gaussian type orbitals and Machine Learning\cite{Brockherde2017}. With Gaussian functions being analytic and infinitely differentiable, any density derivative is analytic, as is any sum thereof, which means that $\tilde\rho$, the lambda-averaged electron density, converges to a finite value.
As shown numerically in the following sections, this sum can be truncated after few terms for iso-electronic alchemical interpolations at fixed nuclei. 
This allows to formulate an energy functional that only depends on the reference electron density $\rho_{\rm r}(\mathbf{r})$
and its perturbations in nuclear charge, or pseudo-potential parameters for that matter\cite{Samuel-JCP2016,Samuel2018bandgaps},
which can be connected to a change in $\lambda$ through repeated use of the chain-rule:
\begin{align}
    \frac{\partial \rho}{\partial \lambda} &= \sum_I \frac{\partial \rho}{\partial Z_I}\frac{\partial Z_I}{\partial\lambda}\\
    \frac{\partial^2\rho}{\partial\lambda^2} &= \sum_I \frac{\partial \rho}{\partial Z_I}\frac{\partial^2Z_I}{\partial\lambda^2} + \sum_J\frac{\partial^2\rho}{\partial Z_I \partial Z_J}\frac{\partial Z_I}{\partial \lambda}\frac{\partial Z_J}{\partial \lambda}
\end{align}
This way, all higher order derivatives contain only the perturbations in nuclear charge of the reference electron density $\rho$. Formally, this approach could be extended to deal also with non-isoelectronic systems through fractional number of electrons. However, due to the known derivative discontinuities with respect to electron number, we would expect much worse performance in practice.

%\section{Results and Discussion}
We find it exciting to note that no quantum calculation is necessary for the target molecule. 
Its specific chemistry enters {\em solely} by virtue of the analytically known terms, the nuclear repulsion energy and $\Delta v(\mathbf{r})$. 
It is therefore obvious to ask if APDFT estimates based on explicitly correlated electron densities can be used to efficiently and reliably estimate the energies and quantum properties of other molecules. 
% \begin{align}
% E\left[\rho(Z_\text{r}), \partial_\lambda^{(1)}\rho(Z_\text{r}), \cdots, \partial_\lambda^{(n)}\rho(Z_\text{r}), \{Z_\text{r}\}, \{Z_\text{t}\}\right]
% \end{align}
While interesting in general, such a functional could be particularly useful for large screening calculations where the total electronic energy of many similar molecules has to be assessed very quickly, since in this case only one self-consistent density is required.

In order to test how fast (and if) above equations converge, we have first estimated the energy of He using alchemical perturbations up to four orders for H$_2$ as a reference system. More specifically, we have used the 
linear annihilation of one proton in H$_2$ (internuclear distance $d =$ 1.0 \AA), 
and simultaneous increase of the nuclear charge in the other atom from 1 to 2. 
Figure~\ref{fig:horder} shows the resulting energy estimate errors as a function of highest order in the Taylor expansion that has been taken into account for this simple two-electron system. 
Regardless of the reference method (HF, LDA, GGA, Hybrid-GGA, CCSD), the error is reduced systematically with higher order terms. 
Due to symmetry in geometry, even expansion orders give symmetric density contributions while odd orders give antisymmetric ones, which means that even orders in the expansion do not contribute due to parity of the integrand. This can be clearly seen in Figure~\ref{fig:horder} where any change is obtained for additional odd expansion terms. 
This example also highlights that vanishing nuclei can be treated without any further adjustment to the method\cite{anatole-jcp2009-2}. 

% discussion figure cobond

\begin{figure}
\includegraphics[width=\columnwidth]{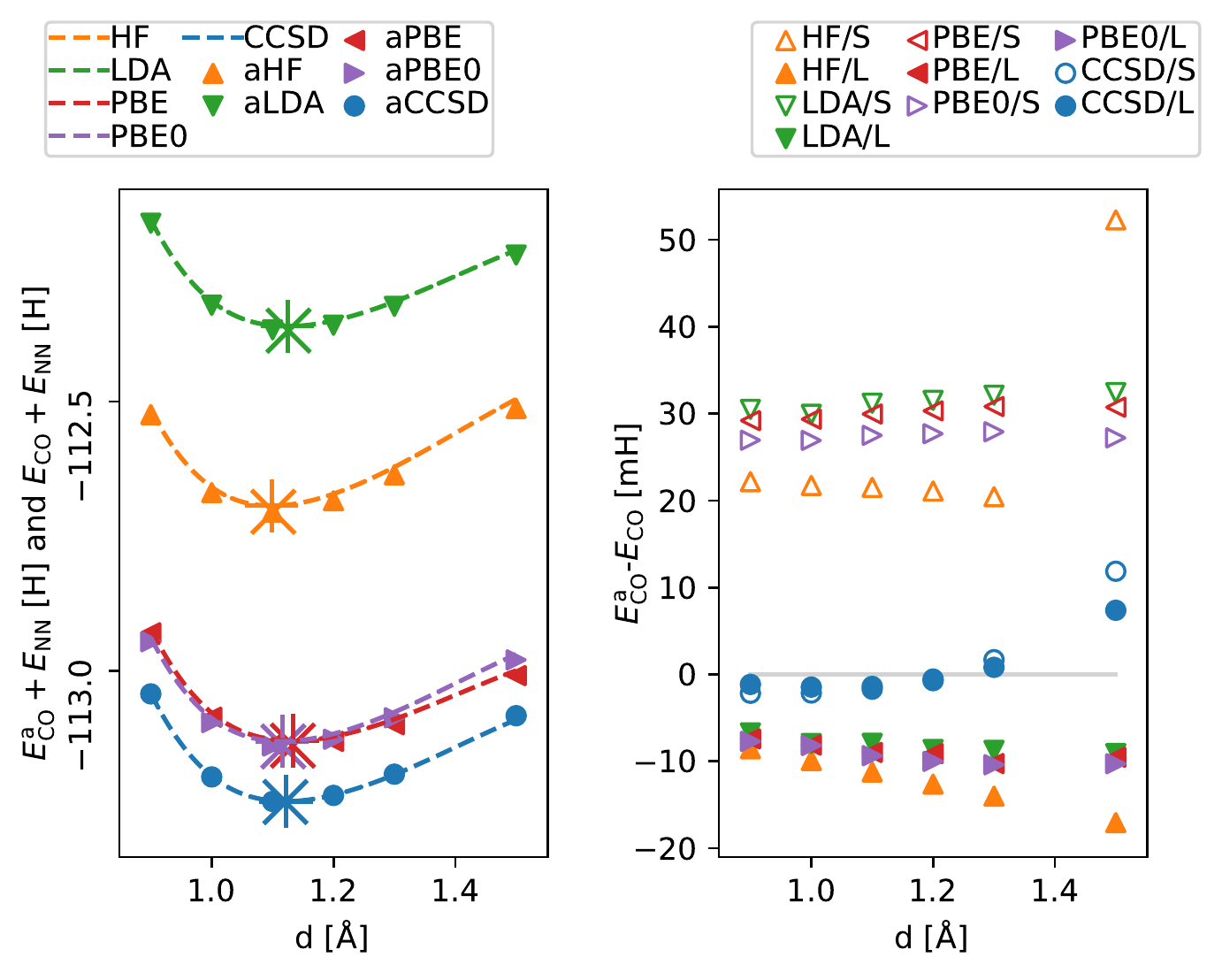}
\caption{Binding potential of CO including the nuclear-nuclear repulsion energy $E_\text{NN}$. Left: Alchemical second order estimates (symbols) obtained from N$_\text{2}$ for various methods (dashed) and def2-TZVP basis set. Plus/cross symbols denote equilibrium bond lengths for target/alchemicy, respectively. 
Right: Error for various methods in small (S=6-31G(d)) and large basis (L=def2-TZVP). Raw data given in the SI.
\label{fig:colot}}
\end{figure}

Going from 2-electron toy model systems, such as H$_2$ and He, to more relevant molecules, 
we have estimated the covalent binding energy of CO perturbing the electron density of N$_2$ up to second
order. Figure~\ref{fig:colot} shows the resulting estimates over a wide range of interatomic distances for various levels
of theory used for the reference calculation. 
It is evident that the proposed method consistently gives numbers close to the actual potential energy curve for any given level of theory that has been used to derive the electron density and its derivatives at the reference molecule. This applies not only to the overall shape but also to the absolute potential energy and highlights that the proposed method approximates the energy of the reference level of theory rather than the true ground state energy. 
Moreover, the location of the minima of the dissociation curves of CO in Figure~\ref{fig:colot} are nearly identical for both the proposed method and the respective reference calculations. While the different level of theory give somewhat different answers for the minimum bond geometry, these differences are conserved when approximating the potential energy surface. 

% Discussion figure colot

Following the potential energy surface over the course of a bond dissociation covers a significant potential energy range. While it is desirable to reproduce the overall shape, systematic accuracy for intermediate distances is needed. This applies both to the range close to the minimum geometry e.g. in the context of geometry optimization and to ranges far from minimum geometry, e.g. in transition states. Figure~\ref{fig:colot} shows the difference between the expected result, i.e. the potential energy of CO with the same basis set and level of theory that has been used for the N$_2$ density, and the true answer, i.e. the energy of the self-consistent CO density. Over a wide range of bond distances the approximate potential energies are accurate to 20-30\,mH for a small 6-31G(d) basis set, while a larger def2-TZVP basis set yields an accuracy of about -10\,mH. This is different for the CCSD densities where -- regardless of basis set -- the accuracy is some 2\,mH for non-dissociative geometries. The stable and systematic error that is exhibited for all levels of theory under investigation shows the consistency of the proposed method. Table I in the SI shows that the energy estimates as obtained from aCCSD are significantly better than the ones obtained from established methods, which suggests to employ few alchemical perturbations at a higher level of theory rather than many lower level calculations. This is with the exception of HF for a small basis set and a bond distance of 1.5\,\AA, where the finite difference scheme we employed introduces numerical artifacts. Note that the finite difference scheme for obtaining the density derivative is by no means a requirement but rather has been used for proof-of-concept work.

In all test cases the use of a larger basis set yields more accurate potential energies. This is in part because the overlap of the atom-centered basis set decreases as the bond length increases, since this offers fewer degrees of freedom for electron density to follow the change in nuclear charges. The major contribution that is visible also in the case of particularly short interatomic distances comes from the finite number of expansion terms. While we have no rigorous proof, the expansion appears to converge faster for a larger basis set. 
Second order perturbation of N$_2$ yields significantly worse results for BF. This is not surprising since the electron density
changes are substantially more dramatic ({\em vide infra}). 
Inclusion of third order terms, however, rectifies the problem and results in improved binding potentials (see SI). 
For comparison, we have also calculated covalent bonding energies in CO and BF using Levy's density averaging approach~\cite{Levy1978}.
Depending on interatomic distance they can respectively deviate by up to $\approx$0.6 and 2.5 Hartree from the CCSD numbers (see SI).

Overall, however, the APDFT results are interesting since they imply that making an alternative investment of compute resources
in high level (for example CCSD in a large basis),
and high order perturbations of reference systems might well enable the screening of an unprecedented
number of alchemically related materials---without sacrificing predictive power. Even considering the increasingly worse scaling of the computational complexity of higher level methods, the combinatorial number of accessible targets scales much faster: for $N$ atoms e.g. in a graphene sheet of which $2n$ are alchemically transformed, the total number of target compounds is
\begin{align}
\sum_{n=1}^N\binom{N}{2n}\binom{2n}{n}\simeq \frac{3^N}{5}
\end{align}
which grows much faster than any of the just polynomially scaling higher level methods (e.g.~$N^q$ with $q=6$ for CCSD).

%Discussion figure ccdesp

\begin{table*}
\begin{ruledtabular}
\begin{tabular}{ll|dddddd||dddddd}
& &\multicolumn{6}{c|}{N$_2 \rightarrow$ CO} & \multicolumn{6}{c}{N$_2 \rightarrow$ BF}\\
Method & $n$ & 
\multicolumn{1}{c}{$|\bm{\mu}|$} & 
\multicolumn{1}{c}{$\delta|\bm{\mu}|$ [\%]} & 
\multicolumn{1}{c}{$Q_\text{xx}$} & 
\multicolumn{1}{c}{$\delta Q_\text{xx}$ [\%]} & 
\multicolumn{1}{c}{$|\mathbf{F}|$} & 
\multicolumn{1}{c|}{$\delta |\mathbf{F}|$ [\%]} & 
\multicolumn{1}{c}{$|\bm{\mu}|$} & 
\multicolumn{1}{c}{$\delta|\bm{\mu}|$ [\%]} & 
\multicolumn{1}{c}{$Q_\text{xx}$} & 
\multicolumn{1}{c}{$\delta Q_\text{xx}$ [\%]} & 
\multicolumn{1}{c}{$|\mathbf{F}|$} & 
\multicolumn{1}{c}{$\delta |\mathbf{F}|$ [\%]} 
\\\hline
\multicolumn{14}{c}{$d = 1.1\,$\AA}\\\hline
CCSD&-&12.54&-&-27.57&-&10.96&-&11.03&-&-25.24&-&9.91&-\\
aCCSD&0&14.55&16.05&-31.37&13.76&12.96&18.22&14.55&31.89&-31.37&24.27&14.58&47.20\\
aCCSD&1&12.48&-0.45&-27.09&-1.77&11.07&0.99&10.41&-5.61&-22.80&-9.65&10.33&4.28\\
aCCSD&2&12.49&-0.41&-27.46&-0.40&10.95&-0.09&10.43&-5.42&-24.31&-3.68&9.80&-1.11\\
aCCSD&3&12.52&-0.12&-27.55&-0.08&10.96&-0.04&10.72&-2.84&-25.01&-0.90&9.85&-0.59\\
aCCSD&4&12.54&-0.01&-27.62&0.18&10.96&-0.04&10.95&-0.72&-26.15&3.61&9.84&-0.64\\\hline
HF&-&12.42&-0.92&-27.43&-0.50&10.82&-1.33&11.07&0.32&-25.74&1.97&9.83&-0.81\\
LDA&-&12.60&0.47&-27.67&0.37&10.91&-0.53&11.09&0.54&-25.09&-0.59&9.92&0.10\\
PBE&-&12.60&0.49&-27.70&0.46&10.85&-1.02&11.10&0.59&-25.17&-0.28&9.88&-0.30\\
PBE0&-&12.54&0.01&-27.59&0.05&10.85&-1.03&11.08&0.45&-25.34&0.40&9.87&-0.41\\\hline\hline
\multicolumn{14}{c}{$d = 1.5\,$\AA}\\\hline
CCSD&-&16.53&-&-47.53&-&6.20&-&13.94&-&-41.81&-&5.55&-\\
aCCSD&0&19.84&20.05&-56.55&18.97&7.32&18.17&19.84&42.37&-56.55&35.25&8.24&48.31\\
aCCSD&1&16.73&1.22&-47.72&0.39&6.25&0.80&13.62&-2.31&-38.89&-7.00&5.81&4.69\\
aCCSD&2&16.64&0.70&-47.34&-0.41&6.24&0.73&13.27&-4.78&-37.37&-10.61&5.80&4.36\\
aCCSD&3&16.10&-2.57&-46.12&-2.98&6.21&0.19&8.95&-35.75&-27.60&-33.99&5.49&-1.09\\\hline
HF&-&16.15&-2.31&-46.79&-1.57&5.96&-3.77&13.93&-0.04&-42.45&1.53&5.43&-2.16\\
LDA&-&16.63&0.60&-47.90&0.78&6.08&-1.82&14.10&1.20&-42.04&0.55&5.52&-0.61\\
PBE&-&16.64&0.66&-47.93&0.84&6.04&-2.49&14.11&1.27&-42.12&0.74&5.49&-1.07\\
PBE0&-&16.47&-0.34&-47.48&-0.10&6.03&-2.69&14.02&0.63&-42.13&0.77&5.48&-1.40\\
\end{tabular}
\end{ruledtabular}
\caption{Dipole moments $\mu$, quadrupole moments $Q_\text{xx}$ and ionic forces $F$ as calculated from the reference CCSD/def2-TZVP densities and the alchemically perturbed CCSD/def2-TZVP densities for CO and BF for two different bond lengths, $d$, and for various expansion orders $n$. All quantities are only electronic, i.e.~without nuclear-nuclear contributions. All data given in a.u., errors $\delta$ relative to CCSD given in percent. \label{tab:ccsd}}
\end{table*}

Having seen that the level of theory for the reference quantum calculations is largely determining the accuracy of the energy predictions, one can wonder how APDFT performs for the prediction of electron density. 
Figure~\ref{fig:ccsdesp} shows densities and electrostatic potentials for CO and BF as calculated from N$_2$. The electron density of CO and the derived electrostatic potential converges quickly. Considering the the dipole moment 
$\mu = \int d\bf{r} \, \rho(\mathbf{r})\mathbf{r}$,
the quadrupole moments 
$Q_{ij} =  \int d\mathbf{r} \, \rho (\mathbf{r})(3r_ir_j-|\mathbf{r}|^2\delta_{ij})$  and the ionic forces
$\mathbf{F}_{\rm I} = Z_{\rm I} \int d\mathbf{r} 
, \rho(\mathbf{r}) (\mathbf{r}-\mathbf{R_\text{I}})/|\mathbf{r}-\mathbf{R_\text{I}}|^3$, shown in Table~\ref{tab:ccsd}: including terms of second order reproduces $\mu$ and $Q$ to about 1\,\%. Since for linear molecules $Q_\text{xx}=Q_\text{yy}$ and $\forall i\neq j: Q_\text{ij}=0$, the electron density also has the expected axial symmetry. Generally, and as one would expect, 
estimates for CO converge more quickly than for BF even though the densities of both molecules are obtained from the same N$_2$ calculations. 
This convergence behavior is expected since the difference in nuclear charges is more moderate for CO than for BF, i.e. the domain to be covered by the Taylor expansion is larger, and, hence, convergence is slower, ultimately limited by the numerical accuracy we can obtain. The limited accuracy of multipole moments as demonstrated in Table~\ref{tab:ccsd} can be understood when taking the electrostatic potential in Figure~\ref{fig:ccsdesp} into account: since the regions more distant to the nuclei converge slower with expansion order, multipole moments are more strongly affected while contributions to ionic forces decay with distance from the nuclei. 

As shown in Table~\ref{tab:ccsd}, HF and the different DFT functionals perform similarly for all cases, highlighting the reliability and black-box nature of the established methods. In comparison to CCSD however, alchemical predictions are often more accurate than their DFT counterparts. The table shows all alchemical predictions up to numerically stable orders, i.e. 3 or 4.

Due to the nature of the density expansion, negative electron densities can arise for intermediate values, e.g. odd orders in the BF case. While negative electron densities are unphysical, at no step the derivation requires the truncated series expansion to be strictly positive. Intermediate deviations from the limit value are generally possible in series expansions. In this case however, it is a sign that higher orders of the expansion should be included, which is illustrated by the improvement from order 1 to 3 in the BF case. $Q_\text{xx}$ and $Q_\text{yy}$ being nearly identical even for high expansion orders is a sign of the numerical stability which conserves the symmetry of the electron density.

%\subsection*{Limitations}
It is important to emphasize the fact that APDFT is not a black-box method which can be applied blindly throughout compositional and configurational space. The choice of reference system, for example, is crucial for the predictive performance. While we have tried to identify and use those reference systems which maximize predictive accuracy in the examples shown above, it should be clear that poor reference choices will lead to poor predictions. 
Furthermore, more fundamental limitations of the method arise from the derivation of the density functional. The density response due to changes in the nuclear charges needs to be continuous. For a rigorous derivation, this response needs to be smooth and the sum building $\tilde\rho$ needs to be uniformly converging. While we are not aware of a formal proof of the latter conditions, one notable case where the density response is sudden would be the H$_2^+$ one-electron system where an infinitesimally small perturbation of the molecular symmetry results in abrupt changes in the entire electron density~\cite{Samuel-JCP2016}. Also, as has been pointed out earlier\cite{Levy1982}, scaling all nuclear charges down, i.e. going from N$_2$ towards and beyond C$_2^{2-}$ can have a discontinuity in the density response when one of the electrons cannot be bound any more. Within a certain radius around the nucleus of a free atom, we show in the SI that the density expansion converges. Another more technical requirement for the density response to be smooth is that the atomic basis functions overlap sufficiently well. This is illustrated by the distance dependency in Figure~\ref{fig:colot}.
Finally, and from a more technical perspective, the proposed method requires the electron density to be mapped on an integration grid. This work uses a Becke-Lebedev\cite{Becke1988,Lebedev1999} grid. To evaluate $\tilde\rho$, the density derivatives w.r.t. nuclear charges need to be available. The employed finite difference scheme however, could be replaced by these derivatives all together. This is desirable since the employed finite difference scheme uses a direct connection of the density between the reference and target molecule, which requires a superposition of all basis sets of the involved atoms. Note that this is only a consequence of the finite difference scheme, and not of the method\cite{Epstein1967}.

%\subsection*{Conclusion}
To conclude, we introduced an orbital free alchemical perturbation  density functional theory (APDFT).  
We have shown that the electron density of target molecules can be constructed by the same reference information in a way that not only forces but also electrostatic potential, dipole moments and quadrupole moments are reproduced. 
The accuracy of such quantum property predictions
converges with perturbation expansion order for all close-by (i.e. $\Delta Z=1$) target systems studied. Similar to forces that allow geometry optimization but are only strictly valid at the reference geometry, the alchemical gradients allow optimization in chemical composition. In the same way how the Hessian can be used for increasing the step length in geometry optimisations, alchemical higher orders are required for target systems farther away in chemical space or with lower density overlap.
Using CCSD reference calculations for N$_2$, APDFT
affords predictions of CO and BF of similar or better quality than PBE0 for energies, forces, and electrostatics already at relatively low perturbation order 3 and 4. 
Since the reference information is identical for all target molecules, estimating quantum properties for any target system comes at negligible additional cost. 

Since only the electron density information is required, APDFT can be applied to any quantum chemistry reference calculation that gives electron densities. We have demonstrated that the accuracy of both energy and density is comparable to the level of theory employed for the one reference calculation if convergence is reached. This means that computational efforts can be shifted from a brute-force screening approach calculating many molecules at intermediate quality to few high quality calculations as a reference for alchemical estimates. 
Depending on accuracy requirements, our results suggest systematic accuracy improvement by inclusion of higher order terms, or, conversely,  coverage of larger regions of chemical space---from one reference perturbation alone.

\begin{acknowledgments}
We acknowledge support by the Swiss National Science foundation (No.~PP00P2\_138932, 407540\_167186 NFP 75 Big Data, 200021\_175747, NCCR MARVEL).
Some calculations were performed at sciCORE (http://scicore.unibas.ch/) scientific computing core facility at University of Basel.
\end{acknowledgments}

\bibliography{main,literatur}
\end{document}